\def\year{2020}\relax
\documentclass[letterpaper]{article} 
\usepackage{aaai20}  
\usepackage{times}  
\usepackage{helvet} 
\usepackage{courier}  
\usepackage[hyphens]{url}  
\usepackage{graphicx} 
\usepackage{graphicx}  
\usepackage{xcolor}
\usepackage{balance}
\graphicspath{{figures/}}

\newlength{\bibitemsep}
\newlength{\bibparskip}
\let\oldthebibliography\thebibliography
\renewcommand\thebibliography[1]{%
  \oldthebibliography{#1}%
  \setlength{\parskip}{\bibitemsep}%
  \setlength{\itemsep}{\bibparskip}%
}

\title{Towards a Policy-as-a-Service Framework to Enable\\
Compliant, Trustworthy AI and HRI Systems in the Wild}

\author{
\Large\textbf{Alexis Morris}\textsuperscript{\rm{1}}, 
\Large \textbf{Hallie Siegel}\textsuperscript{\rm{2}}, \textbf{and Jonathan Kelly}\textsuperscript{\rm{2}} \\ 
\textsuperscript{\rm{1}} Adaptive Context Environments Lab, OCAD University\\
Ontario, Canada, M5T 1W1, amorris@faculty.ocadu.ca\\
\textsuperscript{\rm{2}} University of Toronto Robotics Institute,\\
Ontario, Canada, M5S 3H7, hallie.seigel@utoronto.ca, jkelly@utias.utoronto.ca}

\begin{document}
\maketitle

\vspace*{-4mm}
\begin{abstract}
Building trustworthy autonomous systems is challenging for many reasons beyond simply trying to engineer agents that `always do the right thing.' There is a broader context that is often not considered within AI and HRI: that the problem of trustworthiness is inherently socio-technical and ultimately involves a broad set of complex human factors and multidimensional relationships that can arise between agents, humans, organizations, and even governments and legal institutions, each with their own understanding and definitions of trust. This complexity presents a significant barrier to the development of trustworthy AI and HRI systems---while systems developers may desire to have their systems `always do the right thing,' they generally lack the practical tools and expertise in law, regulation, policy and ethics to ensure this outcome. In this paper, we emphasize the ``fuzzy'' socio-technical aspects of trustworthiness and the need for their careful consideration during both design and deployment. We hope to contribute to the discussion of trustworthy engineering in AI and HRI by i) describing the policy landscape that must be considered when addressing trustworthy computing and the need for usable trust models, ii) highlighting an opportunity for trustworthy-by-design intervention within the systems engineering process, and iii) introducing the concept of a ``policy-as-a-service'' (PaaS) framework that can be readily applied by AI systems engineers to address the fuzzy problem of trust during the development and (eventually) runtime process. We envision that the PaaS approach, which offloads the development of policy design parameters and maintenance of policy standards to policy experts, will enable runtime trust capabilities intelligent systems in the wild.
\end{abstract}

%

\section{Introduction}

Designing trustworthy autonomous systems is challenging in part because the concept of trust derives from human perception and is inherently ``fuzzy'' \cite{adjekum2017trust}. Much of the recent work on trustworthy autonomy (for, e.g., human-robot interaction) has focused on the technical aspects of development and deployment \cite{barattini2019human,zacharaki2020safety,henschel2020social}. However, hardware-software systems engineering is only one part of the problem: trustworthiness is inherently \emph{socio-technical} and comprises a dynamic interplay between the physical, technological, emotional, regulated, and social experience. 
Managing the complexity of trust dynamics inherent in autonomous systems designed to interact with humans is a significant barrier to developers, regulators, and implementers alike. Developers struggle to foresee the ethical and policy implications of their technologies, while regulators struggle to write policy that adequately addresses the moving target of rapid innovation. Meanwhile, organizations that are seeking to implement resource-saving autonomy solutions are uncertain of the safety, legal, and operational risks. To successfully deploy trustworthy autonomous systems ``in the wild,'' however, inputs and constraints from all of these stakeholders must be considered simultaneously. 

System designers and engineers have a tremendous responsibility to consider the impact of human trust on system design and runtime uses. When AI or HRI systems succeed and become more “trustworthy” over time, their human user base learns to transfer increasingly important tasks onto them. When these systems fail, they can have a significant negative impact on the underlying human social, regulatory, and legal networks in which they operate. The fuzzy human concept of trust may be easily misunderstood by stakeholders in the system design and application process, or even misused (potentially maliciously) leading to undesirable outcomes. Hence while a design-time focus and awareness of fuzzy issues like trust is essential, a runtime awareness of the underlying trustworthiness of autonomous systems in the wild is even more significant. And yet the question of how to enable both compliance and trustworthiness in agent systems (whether software agents or embodied robots) remains an open research question that requires dedicated focus within a wide category of  ``social considerations'' \cite{zacharaki2020safety}. In short, the safe diffusion of AI and HRI technologies to industrial, social and consumer applications will require practical translation tools to help stakeholders cooperate and communicate smoothly and efficiently, especially so that ethical, legal, and regulatory issues can be addressed both early in the design cycle, and that compliance can be assured throughout the system’s operational or runtime period.

In this short paper, we argue that the current lack of tools, methodologies and processes to effectively translate ethical, legal, and policy design parameters into engineering practice remains a critical gap in the deployment of trustworthy autonomous systems outside of the laboratory. Currently, regulators and lawmakers tend to identify interesting issues related to trust (such as bias, transparency, accountability, safety, delegated decision-making, etc.) \cite{bejtullahu2018workshop}, and industry leaders raise concerns that autonomous products will simply not find a market if they are perceived as untrustworthy, unethical, impossible to insure, or in breach of the law or existing standards \cite{barattini2019human}. However, engineers struggle to address these indefinite trust issues effectively in their development processes, partly because these issues are not recognized as ``technical,'' partly because there is no clear organizational support (e.g. organizational policies or processes) allocated to addressing them, but primarily because there is no robust technical process for doing so. Without a method or framework for integrating socio-technical trust policies in the AI and HRI development process, there is little hope of addressing trust in operational or runtime terms. To tackle these shortcomings, we propose the need for a policy-as-a-service (PaaS) framework that would incorporate an understanding of trust. By policy-as-a-service, we mean a service-orientated architecture (much like software-as-a-service or robots-as-a-service) that provides just-in-time policy guidance (related to ethics, regulation, compliance, etc.) to non-experts. It is our view that such an approach, which offloads the heavy lifting of policy design and policy standards maintenance to domain specialists, is required to  facilitate the safe transition and diffusion of autonomous technologies from the lab to industrial, social, and consumer applications.

\section{Background Context}

A critical difficulty in designing trustworthy autonomous systems is that trust is a multidimensional ``fuzzy'' concept \cite{adjekum2017trust} that is not measured directly, but is instead measured by proxy. For example, in autonomous vehicles (AVs), trust might be inferred or indirectly measured by the number of accidents, driver interventions, or road infractions per autonomous mile driven. Yet the danger of relying on proxy measures is that they tend to focus our attention too narrowly on what is `measurable,' making us blind to bias, unforeseen risk, and even opportunities. To counterbalance this tendency requires a deliberate broadening of our frame of reference for evaluating trustworthy performance. And yet the large number of different stakeholders involved (developers, operators, regulators, lawmakers, ethicists, end users, and the like), each with their own trust concerns, makes this process unwieldy \cite{barattini2019human,lin2017robot}. Furthermore, some aspects of trust (for example, social trust) are poorly understood and may simply not be measurable in the traditional sense. How to safely deploy autonomous systems in the wild thus presents a classic ``wicked problem''---an intractable problem that is impossible to ``solve'' given multidimensional, dynamic interactions involving different stakeholders with conflicting priorities and needs \cite{bejtullahu2018workshop}. As a result, and unsurprisingly, much of the discussion on how to regulate autonomous technologies has, to date, been unproductive. Although there are increasing calls at the international level to include policy, ethics, and values in the technology design process (World Economic Forum 2018 \cite{philbeck2018values}; United Nations 2018 \cite{prestes2019overview,general2019age}), few practical opportunities currently exist to do so. Further, as policymakers and engineers come together, the emphasis can often tend towards debate rather than collaboration and the development of shared values \cite{philbeck2018values}.

A scientific response to the wicked problem of trust in autonomous systems can easily conclude that it is out-of-scope because there are too many factors that cannot be controlled, let alone understood \cite{lin2017robot}. But leaving the question solely to industry, or social scientists, policymakers, ethicists, and lawmakers is unlikely to address the issue either, because there are dynamic interdependencies between all these disciplines that cannot be addressed in isolation \cite{philbeck2018values}. Isolated solutions to one problem therefore risk generating unintended consequences in another area. In short, working in isolation leads to blind spots that undermine the resilience of entire socio-technical AI and HRI systems. This points to the need for a system-of-systems approach to ensure the resilience and robustness \cite{philbeck2018values}. Moreover, while trust is central to the successful adoption of autonomous technologies, it must be transformed from a fuzzy human factor into a practical design consideration if we are to widely deploy AI and HRI systems for societal benefit.

A related problem is that definitions of ``trust'' and ``trustworthiness'' differ from discipline to discipline (e.g., they may be variously interpreted as risk, predictability, robustness, safety, or accountability), creating confusion and miscommunication between stakeholders. Furthermore, the trustworthiness of an autonomous system is experienced in different ways by different stakeholders at different times, and is dependent on many overlapping factors such as prior experience, operational context, the regulatory environment, social experience, and personal intuition. In addition, with governments and industry in other regions investing heavily, the autonomous systems sector is both highly competitive and evolving quickly, making it a challenge to anticipate future capabilities and their associated risks \cite{zacharaki2020safety}.

\begin{figure*}[t]
    \centering
    \includegraphics[clip,trim=0 6pt 0 0,width=0.80\linewidth]{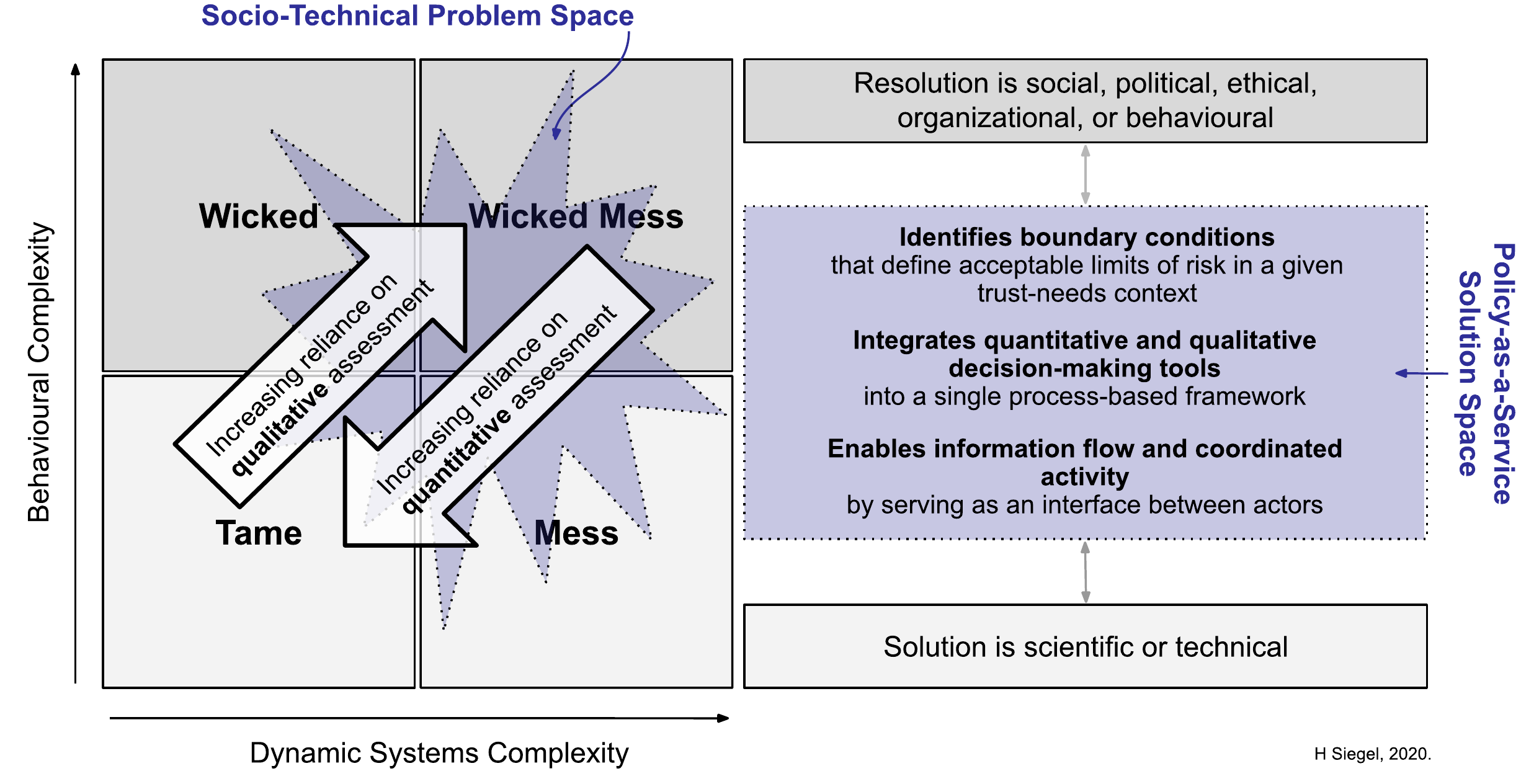}
    \caption{The Socio-Technical Problem Space (left) and the associated Policy-as-a-Service solution space (right), adapted from \cite{hancock2003tame}.}
    \label{fig:paas}
    \vspace{-2mm}
\end{figure*}

\section{Towards a Policy-as-a-Service Framework}

The wicked nature of the problem of deploying trustworthy AI and HRI systems demands a new, integrated view of trust dynamics that cuts across traditional boundaries between design, policy, ethics, and other areas. If we are to muddle our way towards innovating policy (specifically) in this space, we must acknowledge that i) trust plays a critical role throughout the entire innovation cycle, and is influenced by many actors; ii) there are many delays in the current system’s underlying innovation/regulation feedback loop (caused by the time it takes to write new laws, for example), and iii) trust is impacted by not just the performance gap, but also the experience gap between the technology’s expected performance and that experienced by the user.

As noted previously, one of the main gaps for the safe deployment of autonomous systems (whether state-deployed or otherwise) is the lack of tools, methodologies and processes that effectively translate ethical, legal and policy issues into engineering practice. 
Engineers typically do not have the expertise to address these issues effectively in their organizations and in their development processes. At the same time, the issues resonate with policymakers and the public, and increasingly, with business leaders who understand that their products will simply not find a market if they are perceived as untrustworthy, unethical, impossible to insure, or in breach of the law \cite{philbeck2018values}. In short, there is a clear need for practical translation tools to help stakeholders cooperate and communicate smoothly and efficiently, so that ethical, legal, and regulatory issues can be addressed earlier in the design cycle.

\subsection{Policy-as-a-Service and Complexity}

The above exploration of the complexity of the trust problem in autonomous systems highlights the need for a PaaS framework. Such a framework would fill a gap and act as a step toward a holistic, human-centered, `techno-socio-legal' approach to mapping trust dynamics (with robotics as an example), for future development of a flexible architecture for enhancing the trustworthiness of autonomous systems. The framework must support metrics for evaluating trustworthiness and also provide meaningful human control, that is, true operational oversight, of autonomous systems. This encompasses not just the end-user interactions, but law, regulatory, and organizational policy as well. The overarching goal would be to enable future organizations to measure, evaluate, and predict the trustworthiness of the autonomous systems that they deploy, for the purpose of safely assigning decision-making authority to either AI or human agents.

Figure 1 illustrates the inherent complexity of the socio-technical problem space (purple starburst) and the required flexibility of the PaaS solutions space (purple dashed rectangle), mapped onto Hancock and Holt’s Risk/Problem-Type Relationship Matrix \cite{hancock2003tame}. The Socio-Technical Problem Space (left side) shows how human and/or organizational behaviour can interact with autonomous system dynamics to create varying degrees of socio-technical complexity. Simple socio-technical interactions with limited behavioural complexity and technical dynamics would be classified as relatively ``Tame'' and could be modelled and assessed by primarily quantitative and rules-based means. As behavioural and system dynamics become increasingly complex, characterizing and predicting socio-technical interactions becomes an intractable ``Wicked Problem'' that cannot be solved quantitatively, but must instead rely increasingly on qualitative or principles-based approaches. The Policy-as-a-Service Solution Space (right) requires a framework that can integrate qualitative and quantitative risk assessment tools, allowing organizations to i) map the boundary conditions for which aspects of law, ethics, and policy can be safely and reasonably quantified and parameterized for use in autonomous decision-making, and ii) determine which aspects ought to be reserved for qualitative (human) decision-making, in order to ensure meaningful human control. 

\subsection{Policy-as-a-Service Design Considerations}

The design of a PaaS framework is certainly nontrivial and will require a series of collaborative, iterative research and prototyping activities aimed at progressively integrating several interdependent research strands: policy (to focus on the legal, ethical interface), communications (to focus on the sensor network, privacy, and security interface), systems engineering (to focus on the software and hardware interface), learning and prediction models (to focus on the algorithmic interface), and organizational learning (to focus on inter and intra-organizational communication interfaces). Importantly, design challenges generally occur at interfaces.

Towards our goal, we have identified the following activities to be considered by interdisciplinary groups when developing a PaaS framework for autonomous systems. This list is not exhaustive, but highlights some of the key issues involved:
\begin{enumerate}
	\item The development of applied sub-frameworks towards understanding and anticipating the complexity of trust dynamics in a multidimensional space (which could include defense, law, ethics, policy, industry, engineering, foresight, and user-centered design).
	\item The clarification of language regarding what trust means from different stakeholder perspectives, towards a shared understanding and common vision of trust.
	\item The design of applied use cases, collaboratively anticipating and modelling the complexity of trust dynamics involved when deploying mobile autonomous systems (such as autonomous vehicles).
	\item The establishment of a `community of practice' that supports the development of trustworthy autonomous systems and the investigation of standards.
	\item The identification of critical interactions between stakeholders that promote safety, resilience, robustness, and transparency.
	\item The sharing of human-centered design principles and processes across the members of the developer community.
	\item The involvement of policy experts in the development process and the data outputs toward policy-based anticipation and response to rapidly evolving technology.
	\item The design of administrative and communication processes that incorporate iterative stakeholder feedback into the development process.
	\item The collaborative development and prototyping of a flexible, PaaS support architecture for autonomous systems.
	\item The revealing of structural interdependencies by collectively defining the trust model toward applied uses.
	\item The development and iterative refinement of shared design criteria and standards for the policy-as-a-service architectural framework and prototype systems.
	\item The identification of (if any) ethical, legal, and organizational policy features can be effectively parameterized and incorporated by developers into the behaviour of autonomous systems, and identifying the inherent risks of doing so.
	\item The iterative development and testing of metrics for evaluating the policy-as-a-service framework, and the trustworthiness of the autonomous systems that use it.
	\item The complete testing of the policy-as-a-service framework in the context of operation of a real-world autonomous system scenario.
\end{enumerate}

These joint activities are steps on the path to enabling researchers and developers of these systems to deepen their understanding of the trust dynamics involved with autonomous systems, while recognizing the variety of stakeholder perspectives in their respective fields.

At various iterations, specific collaborative research outputs of the design process would include increasingly sophisticated prototypes for: i) a multidisciplinary trust model and trust taxonomy; ii) a policy-as-a-service, including: design criteria, framework, and evaluation metrics; iii) policy-learning algorithms, and iv) trust-based education and design toolkits for practitioners and deployers of autonomous systems. As the various prototypes evolve out of this iterative process, investigators and collaborators would shift toward rigorous evaluation through simulation, demonstrations and/or user-testing within the stakeholder's disciplinary networks. This approach to designing a PaaS framework for runtime uses based closely on stakeholder perspectives and participation is a methodology that would allow for a cross-section of disciplines to engage with the autonomous systems design process, while at the same time promoting values-based, user-centered design to the autonomous systems community. These steps will take time and effort, certainly. Despite this, we assert that without PaaS-type frameworks, the successful deployment of large-scale AI and HRI systems (e.g., massive fleets of self-driving vehicles) will remain elusive.

\section{Discussion and Summary}

As autonomous systems become more capable, they are expected to perform their tasks in ways that can often stress their foundational designs, particularly as users of these systems gain confidence in these devices across a variety of situations. This factor of trust and reliability in system designs often can be impacted at runtime, whenever a system fails to perform as expected, causing inconvenience, injury, or even death. Such failures can be considered as a breach of trust, particularly when they are deployed or sanctioned by an authority. To prevent trust breaches, there is a need for embedding trust-related intelligence into autonomous systems, and the general trustworthiness of these systems must be explored. Our position is that, to make any meaningful progress towards ensuring trustworthy autonomous systems, trust must be transformed from a fuzzy human factor into a practical design consideration. We have highlighted the need for a flexible, policy-as-a-service framework and supporting architecture that i) broadens proxy definitions of trust to encompass physical, legal, social, and policy performance metrics; ii) is iteratively co-designed to enable cooperation between the many stakeholders developing and impacted by autonomous systems; iii) has parameterized design criteria so that legal, ethical and organizational policy can be made tangible to hardware and software developers; and iv) enables operators to measure, evaluate, and predict the trustworthiness of the autonomous systems they deploy in the appropriate context. The next step is to begin the (difficult but highly valuable) process of building a prototype PaaS framework.

\section*{Acknowledgements}

The authors acknowledge the impact of fruitful discussions with knowledge experts within the Trust-by-Design Micronet initiative (2019), particularly Jason Millar, Jutta Treviranus, Ian Kerr, Frauke Zeller, AJung Moon, Ken Doyle, and Raheena Dahya. Alexis Morris and Jonathan Kelly gratefully acknowledge support from the Tri-Council of Canada under the Canada Research Chairs program. Jonathan Kelly is a Faculty Affiliate at the Vector Institute for Artificial Intelligence.

\balance
\setlength{\bibitemsep}{0.4\baselineskip plus .05\baselineskip minus .05\baselineskip}
\bibliographystyle{aaai}
\bibliography{biblio}

\end{document}